\documentclass[sigconf,screen]{acmart}

\copyrightyear{2020}
\acmYear{2020}
\setcopyright{acmcopyright}
\acmConference[ESEC/FSE '20]{Proceedings of the 28th ACM Joint European Software Engineering Conference and Symposium on the Foundations of Software Engineering}{November 8--13, 2020}{Virtual Event, USA}
\acmBooktitle{Proceedings of the 28th ACM Joint European Software Engineering Conference and Symposium on the Foundations of Software Engineering (ESEC/FSE '20), November 8--13, 2020, Virtual Event, USA}
\acmPrice{15.00}
\acmDOI{10.1145/3368089.3409694}
\acmISBN{978-1-4503-7043-1/20/11}

\usepackage{amsmath}
\usepackage{amssymb}
\usepackage{mathtools}
\usepackage{makecell}
\usepackage{graphicx}
\usepackage{booktabs}
\usepackage{listings}
\usepackage{xcolor}
\usepackage{tabularx}
\usepackage{filecontents} 
\usepackage{enumitem}
\usepackage[labelfont=bf,font=small,skip=0pt]{caption}
\usepackage[labelfont=bf,font=small,skip=0pt]{subcaption}
\usepackage[T1]{fontenc}
\usepackage[utf8]{inputenc}
\usepackage[capitalise,noabbrev]{cleveref}
\usepackage{balance}

\usepackage{pifont} 
\usepackage{microtype} 

\newlength{\thickarrayrulewidth}
\setlist[description]{leftmargin=0pt,labelindent=0pt}

\definecolor{olivegreen}{rgb}{0.33, 0.42, 0.18}
\definecolor{darklava}{rgb}{0.28, 0.24, 0.2}
\definecolor{lightgray}{gray}{0.4}
\definecolor{darkgray}{gray}{0.2}
\definecolor{shadegray}{gray}{0.9}

\newcommand*{\tblhead}[1]{{\small\rotatebox{65}{\makecell[c]{#1}}}}
\newcommand*{\tblentry}[1]{{\small#1}}
\newcommand*{\allcap}[1]{{\small#1}}

\renewcommand{\arraystretch}{1.1}
\newcommand*{\etal}{et~al.}
\newcommand*{\bcov}{{\textsf{bcov}}}
\newcommand*{\plusinfty}{\text{+\raisebox{-0.25ex}{\Large$\infty$}}}

\newcommand{\HiLight}{\makebox[0pt][l]{\color{shadegray}\rule[-2pt]{0.9\columnwidth}{1em}}}

\lstdefinelanguage{disasm-x64}{
    classoffset = 1,
    morekeywords = {lea,mov,call,cmp,jae,movzx,ja,movsxd,add,jmp,sete,pop,ret,nop,inc,test,jne,jns,sub,adrp,strb},
    keywordstyle = \color{olivegreen},
    morecomment = [l]{;}
}

\lstdefinestyle{custom-x64}{
    belowcaptionskip=0pt,
    abovecaptionskip=0pt,
    framerule=0.6pt,
    basewidth=0.55em,
    xleftmargin=1pt,
    framexleftmargin=1pt,    
    language=disasm-x64,
    basicstyle=\small\linespread{0.5}\ttfamily,
    commentstyle=\color{lightgray},
    breaklines=true,
    escapechar=|,
    showstringspaces=false,
    showspaces=false
}

\lstdefinestyle{custom-cpp}{
    belowcaptionskip=0pt,
    abovecaptionskip=0pt,
    breaklines=true,
    xleftmargin=0pt,
    framexleftmargin=0pt,    
    language=C++,
    showstringspaces=false,
    basicstyle=\small\linespread{0.5}\ttfamily,
    keywordstyle=\color{olivegreen},
    commentstyle=\color{lightgray},
    identifierstyle=\color{black},
    stringstyle=\color{orange},
}

\begin{document}

\date{}

\title{Efficient Binary-Level Coverage Analysis}

\author{M. Ammar Ben Khadra}
\affiliation{%
	\email{stoffel@eit.uni-kl.de}
	\institution{Technische Universit{\"a}t Kaiserslautern}
    \country{Germany}
}
\email{khadra@eit.uni-kl.de}

\author{Dominik Stoffel}
\affiliation{%
	\institution{Technische Universit{\"a}t Kaiserslautern}
    \country{Germany}
}
\email{stoffel@eit.uni-kl.de}

\author{Wolfgang Kunz}
\affiliation{%
	\institution{Technische Universit{\"a}t Kaiserslautern}
    \country{Germany}
}
\email{kunz@eit.uni-kl.de}


\begin{abstract} 
Code coverage analysis plays an important role in the software testing process. 
More recently, the remarkable effectiveness of coverage feedback has triggered a broad interest in feedback-guided fuzzing.
In this work, we introduce {\bcov}, a tool for binary-level coverage analysis.
Our tool statically instruments x86-64 binaries in the ELF format without compiler support.
We implement several techniques to improve efficiency and scale to large real-world software.
First, we bring Agrawal's probe pruning technique to binary-level instrumentation and effectively leverage its superblocks to reduce overhead.
Second, we introduce \textit{sliced microexecution}, a robust technique for jump table analysis which improves CFG precision and enables us to instrument jump table entries.
Additionally, smaller instructions in \mbox{x86-64} pose a challenge for inserting detours.
To address this challenge, we aggressively exploit padding bytes and systematically host detours in neighboring basic blocks.

We evaluate {\bcov} on a corpus of 95 binaries compiled from eight popular and well-tested packages like FFmpeg and LLVM.
Two instrumentation policies, with different edge-level precision, are used to patch all functions in this corpus - over 1.6 million functions.
Our precise policy has average performance and memory overheads of 14\% and 22\% respectively.
Instrumented binaries do not introduce any test regressions.
The reported coverage is highly accurate with an average F-score of 99.86\%.
Finally, our jump table analysis is comparable to that of \textsf{IDA Pro} on \textsf{gcc} binaries and outperforms it on \textsf{clang} binaries.

\end{abstract}



\maketitle

\section{Introduction}
\label{sec:introduction}

Code coverage analysis is commonly used throughout the software testing process~\cite{Ammann2016}.
Structural coverage metrics such as statement and branch coverage can inspire confidence in a program under test (PUT), or at least identify untested code~\cite{Inozemtseva2014,Gopinath2014}.
Additionally, coverage analysis has demonstrated its usefulness in test suite reduction~\cite{Yoo2010}, fault localization~\cite{Pearson2017}, and detection of compiler bugs~\cite{Le2014}.
Moreover, certain coverage requirements are mandated by the standards in safety-critical domains~\cite{DO178C,ISO26262:2018}.

In recent years, feedback-guided fuzzing has emerged as a successful method for automatically discovering software bugs and security vulnerabilities~\cite{VUzzerRawat2017,kAFL:Schumilo2017,QSYMYun2018,Bohme2016a}. 
Notably, AFL~\cite{ZalewskiAFLWhitePaper} has pioneered the usage of code overage as a generic and effective feedback signal.
This success inspired a fuzzing ``renaissance'' and helped move fuzzing to  industrial-scale adoption like in Google's \mbox{OSS-Fuzz}~\cite{OSSFUZZ}.

In this work, we introduce {\bcov}, a tool for binary-level coverage analysis using static instrumentation.
{\bcov} works directly on x86-64 binaries in the ELF format without compiler support. 
It implements a trampoline-based approach where it inserts \textit{probes} in targeted locations to track \textit{basic block} coverage. 
Each probe consists of a \textit{detour} that diverts control flow to a designated \textit{trampoline}. 
The latter updates coverage data using a single pc-relative \texttt{mov} instruction, potentially executes relocated instructions, and then restores control flow to its original state.
Making this scheme to work efficiently and transparently on large and well-tested C and C++ programs required addressing several challenges:

\textbf{Probe pruning}~(\S\ref{sec:probe-pruning}). 
Instrumenting all basic blocks (BBs) can be inefficient, or even impossible, in x86-64 ISA due to its instruction-size variability.
We adopt the probe pruning technique proposed by Agrawal~\cite{Agrawal1994} where dominator relationships between BBs are used to group them in superblocks (SBs).
SBs are  arranged in a superblock dominator graph.
Covering a single BB implies that all BBs in the same SB are also covered, in addition to SBs dominating the current SB.
This allows us to significantly reduce the instrumentation overhead and size of coverage data.

\textbf{Precise CFG analysis}~(\S\ref{sec:controlflow}). 
Imprecision in the recovered control flow graph (CFG) can cause false positives in the reported coverage. 
It can also cause instrumentation errors which lead to crashes in a PUT.
To address this challenge, we propose \textit{sliced microexecution}, a precise and robust technique for jump table analysis. 
Also, we implement a non-return analysis that eliminates spurious CFG edges after non-return calls.
Our experiments show that {\bcov} can outperform IDA Pro, the leading industry disassembler.

\textbf{Static instrumentation}~(\S\ref{sec:instrumentation}). 
Given a set of BBs in an SB, we need to choose the best BB to probe based on the expected overhead of restoring control flow. 
We make this choice using a classification of BBs in \mbox{x86-64} into 9 types.
Also, some BBs can be too short to insert a detour. Their size is less than 5 bytes. 
We address this challenge by (1) aggressively exploiting padding bytes, (2) instrumenting jump table entries, and (3) introducing a greedy strategy for \textit{detour hosting} where a larger BB can host the detour of a neighboring short BB.
Combining these techniques with probe pruning enables tracking coverage of virtually all BBs.

\subsection{Design Overview}

\begin{figure}[t!]
    \centering
    \includegraphics[clip, trim=5cm 5.8cm 10cm 4.8cm, width=0.35\textwidth]{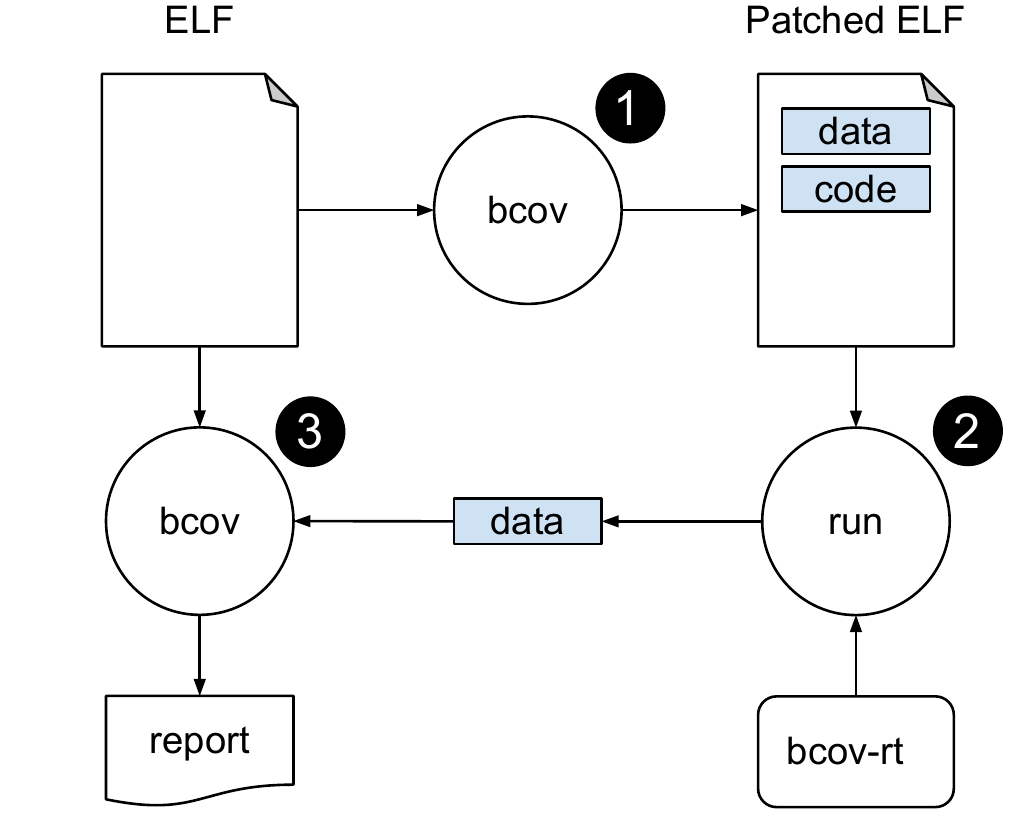}
    \caption[workflow]{The general workflow of {\bcov}. 
        A binary is patched with extra code segment (trampolines) and data segment (coverage data). 
        Our \mbox{\textsf{bcov-rt}} library dumps the data segment at run-time. 
        In our prototype, reporting coverage requires re-analyzing the binary.}
    \Description{The general workflow of bcov}
    \label{fig:bcov-work-flow}
\end{figure}

Figure~\ref{fig:bcov-work-flow} depicts the workflow of {\bcov}.
Given an ELF module as input, {\bcov} first analyzes module-level artifacts, such as the call graph, before moving to function-level analyses to build the CFG and dominator graphs.
Then, {\bcov} will choose appropriate probe locations and estimate the required code and data sizes depending on the \textit{instrumentation policy} chosen by the user.
Our prototype supports two instrumentation policies.
The first is a \textit{complete} coverage policy where for \textit{any} test input it is possible to precisely identify covered BBs. 
The second one is a \textit{heuristic} coverage policy where we probe only the leaf SBs in the superblock dominator graph.
Running a test suite that covers \textit{all} leaf SBs implies that 100\% code coverage is reached. 
We refer to these policies as \textit{any-node} and \textit{leaf-node} policies respectively.
On average, the any-node policy probes 46\% of BBs compared to 30\% in the leaf-node policy.
Average performance overheads are 14\% and 8\% respectively.

The patching phase can start after completing the previous analysis phase.
Here, {\bcov} first extends the ELF module by allocating two loadable segments: a code segment where trampolines are written and a data segment for storing coverage data.
Then, {\bcov} iterates over all probes identified by the chosen instrumentation policy.
Each probe represents a single SB. 
Generally, patching a probe requires inserting a detour targeting its corresponding trampoline.
The detour can be a pc-relative \texttt{jmp} or \texttt{call} instruction.
The trampoline first updates coverage data and then restores control flow to its state in the original module as depicted in Figure~\ref{fig:patch-example}.

The data segment has a simple format consisting of a small header and a byte array that is initialized to zeros. 
Setting a byte to one indicates that its corresponding SB is covered. 
It is trivial to compress this data on disk as only the LSB of each byte is used.
For example, this enables storing complete coverage data of \texttt{llc} (LLVM backend) in 65KB only.~\footnote{The binary has around $1\times10^6$ BBs which contain more than $4\times10^6$ instructions.}
Our data format also enables merging coverage data of multiple tests using a simple bitwise OR operation.

Dumping coverage data requires linking against \textsf{bcov-rt}, our small runtime library.
Alternatively, \textsf{bcov-rt} can be injected using the 
{\allcap{LD\_PRELOAD} mechanism to avoid modifying the build system. 
Coverage data can be dumped on process shutdown or upon receiving a user signal.
The latter enables \textit{online} coverage tracking of long-running processes. 
Note that the data segment starts with a magic number which allows \textsf{bcov-rt} to identify it.

\begin{figure}[t!]
\centering
\vspace{-4pt}    
\begin{subfigure}[t]{0.23\textwidth}
	\begin{lstlisting}[style=custom-x64]
36b62: cmp  eax,0x140
36b67: sete al
36b6a: jmp  36bce
	\end{lstlisting}
	\vspace{-3pt}
	\caption{original code}
	\label{fig:2:original}
\end{subfigure}
\hfill
\begin{subfigure}[t]{0.23\textwidth}
	\begin{lstlisting}[style=custom-x64]
36b62: cmp  eax,0x140
36b67: jmp  6002b8
|
	\end{lstlisting}
	\vspace{-3pt}
	\caption{patched code}
	\label{fig:2:patch}
\end{subfigure}
\hfill
\begin{subfigure}[t]{0.38\textwidth}
	\vspace{-3pt}     
	\begin{lstlisting}[style=custom-x64]
6002b8: mov  BYTE PTR [rip+0xadd88],1
6002bf: sete al
6002c2: jmp  0x36bce
	\end{lstlisting}
	\vspace{-2pt}
	\caption{trampoline}
	\label{fig:2:trampoline}
\end{subfigure}

\caption{{\bcov} patching example. 
	(a) instruction at \textsf{0x36b67} must be relocated as the size of jump at \textsf{0x36b6a} is only two bytes. 
	(b)~relocated instructions are replaced with a 5 byte detour at \textsf{0x36b67}. 
	(c)~coverage update happens at \textsf{0x6002b8}. 
	Control flow is then restored after executing the relocated instruction at \textsf{0x6002bf}.}
    \Description{bcov patching example}
	\label{fig:patch-example}
\end{figure}

This design makes \textsf{bcov} achieve three main goals, namely, transparency, performance, and flexibility.
Program transparency is achieved by not modifying program stack, heap, nor any general-purpose register. 
Also, coverage update requires a single pc-relative \texttt{mov} instruction which has a modest performance overhead.
Finally, \textsf{bcov} works directly on the binary without compiler support and largely without changes to the build system. 
This enables users to flexibly adapt their instrumentation policy without recompilation.


To summarize, we make the following key contributions:
\begin{itemize}
    \item We are the first to bring Agrawal's probe pruning technique to binary-level instrumentation.
    We show that its superblocks can be effectively leveraged to optimize probe selection and reduce coverage data.
    
    \item We introduce \textit{sliced microexecution}, a robust method for jump table analysis.
    It significantly improves CFG precision and allows us to instrument jump table entries.
    
    \item We significantly push the state of the art in trampoline-based static instrumentation and show that it can be used to track code coverage efficiently and transparently.
         
\end{itemize}

We implemented our contributions in the tool \textsf{bcov}, which we make publicly available: 
\url{https://doi.org/10.5281/zenodo.3876047}

We extensively experimented with \bcov. 
In this respect, we selected 8 popular and well-tested subjects such as \texttt{ffmpeg} and \texttt{llc}.
We compiled them using 4 recent major versions of \textsf{gcc} and \textsf{clang} at 3 different optimization levels each.
In total, we used \textsf{bcov} to instrument 95 binaries and more than 1.6 million functions.
Instrumented binaries did not introduce any test regressions.


\section{Motivation}
\label{sec:motivation}

There is a plethora of tools dedicated to coverage analysis.
They vary widely in terms of goals and features.
Therefore, we motivate the need for our approach via a comparison with a representative set of popular tools. 
Our discussion is based on Table~\ref{tab:tool-comparison}.

We start with source-level tools supported in \textsf{gcc} and \textsf{clang}, which are \textsf{gcov} and  \textsf{llvm-cov} respectively.
Both track similar artifacts such as statement coverage.
The key difference is in the performance of instrumented binaries.
\textsf{gcov} can not accurately track code coverage in optimized builds.
In comparison, \textsf{llvm-cov} features a custom mapping format embedded in LLVM's intermediate representation (IR).
This allows it to cope better with compiler optimizations.
Also, this mapping format tracks source code regions with better precision compared to \textsf{gcov}.

The ability of a binary-level tool such as \textsf{bcov} to report source-level artifacts is limited by the binary-to-source mapping available.
Off-the-shelf debug information can be used to report statement coverage - the most important artifact in practice~\cite{IvankovicFSE19, Gopinath2014}.
In this setting, \textsf{bcov} offers several advantages including:
(1)~detailed view of individual branch decisions regardless of the optimization level,
(2)~precise handling of non-local control flow such as \texttt{longjmp} and C++ exception handling, and
(3)~flexibility in instrumenting only a selected set of functions, e.g., the ones affected by recent changes, which is important for the efficiency of continuous testing~\cite{IvankovicFSE19}.

\begin{table}[t!]
    \centering
    \renewcommand{\arraystretch}{1.0}
    \setlength\tabcolsep{2.5pt}
    \caption{A comparison with representative coverage analysis tools. 
    Compiler-dependent tools require modifying the build system and recompilation which limits flexibility. 
    The usability of binary-level tools in the testing workflow is limited. 
    In contrast, {\bcov} only requires replacing a binary with an instrumented version.}
    \label{tab:tool-comparison}    
    \begin{tabularx}{\columnwidth}{@{}lcccccc@{}}
        \\ 
        &
        \tblhead{Level} &
        \tblhead{Coverage\\goal} &
        \tblhead{Compiler\\independence} &
        \tblhead{Performance\\overhead} &
        \tblhead{Flexibility} &
        \tblhead{Usability} \\
        \toprule
        \textsf{gcov}        & \tblentry{source}    & \tblentry{complete} 	& \ding{53} & \ding{53}	  & \ding{53}	   & \ding{51}       \\
        \textsf{llvm-cov}    & \tblentry{source}    & \tblentry{complete} 	& \ding{53} & \ding{51}   & \ding{53}    & \ding{51}             \\
        \textsf{sancov}      & \tblentry{IR}        & \tblentry{heuristic} 	& \ding{53} & n/a         & \ding{53}    & \ding{51}             \\
        \textsf{Intel PT}    & \tblentry{binary}    & \tblentry{heuristic}  & \ding{51} & \ding{51}   & \ding{53}    & \ding{53}       \\
        \textsf{drcov}         & \tblentry{binary}    & \tblentry{both} 	    & \ding{51} & \ding{53}   & \ding{51}    & \ding{53}     \\
        {\bcov}              & \tblentry{binary}    & \tblentry{both}	      & \ding{51} & \ding{51}   & \ding{51}    & \ding{51}    \\
        \bottomrule
    \end{tabularx}
\end{table}

The recent fuzzing renaissance has motivated the need to improve efficiency by heuristically tracking coverage.
SanitizerCoverage (\textsf{sancov})~\cite{SanitizeCoverageURL}  is a pass built into LLVM which supports collecting various types of feedback signals including basic block coverage.
It is used in prominent fuzzers like LibFuzzer~\cite{LibFuzzerWebsite} and Honggfuzz~\cite{SwieckiHonggfuzz}.
The performance overhead of \textsf{sancov} is not directly measurable as the usage model varies significantly between \textsf{sancov} users.
Also, \textsf{sancov} is tightly coupled with LLVM sanitizers (e.g., ASan) which add varying overhead.
Extending {\bcov} with additional feedback signals, similar to \textsf{sancov}, is an interesting future work.

Hardware instruction tracing mechanisms, like Intel\textsuperscript{\textregistered}~PT (IPT), can also be used for coverage analysis.
However, IPT can dump gigabytes of compressed trace data within seconds which can be inefficient to store and post-process.
In our experiments, IPT dumped 6.5\,GB trace data for a \textsf{libxerces} test that lasted only 5 seconds. 
Post-processing and deduplication took more than 3 hours.
In comparison, our tool can produce an accurate coverage report for the same test after processing a 53\,KB dump in a few seconds.
Schumilo {\etal}~\cite{kAFL:Schumilo2017} propose to heuristically summarize IPT data on the fly and thus avoid storing the complete trace.

Dynamic binary instrumentation (DBI) tools can report binary-level coverage using dedicated clients (plug-ins) like \textsf{drcov}.
DBI tools act as a process virtual machine that JIT-emits instructions to a designated code cache.
This process is complex and may break binaries. 
Moreover, JIT optimizations add overhead to the whole program even if we are only interested in a selected part such as a shared library.
Our evaluation includes a comparison with the popular DBI tools  \textsf{Pin}~\cite{IntelPinWeb} and \textsf{DynamoRIO}~\cite{DynamoRIOWeb}.



\section{Probe Pruning}
\label{sec:probe-pruning}

We provide here the necessary background on the probe pruning techniques implemented in {\bcov} based on Agrawal~\cite{Agrawal1994}. 
The original work considered source-level pruning but only for C programs.

Given a function $F$ with a set of basic blocks $B$ connected in a CFG. 
The straightforward way to obtain complete coverage data is to probe every basic block $bb \in B$. 
However, it is possible to significantly reduce the number of required probes by computing \textit{dominance} relationships between basic blocks in a CFG. 
We say that $bb_i$ predominates $bb_j$, $bb_i\xrightharpoondown{pre}bb_j $, iff every path from function entry ($EN$) to $bb_j$ goes through $bb_i$.
Similarly, $bb_i$ postdominates $bb_j$, $bb_i\xrightharpoondown{post}bb_j $, iff every path from $bb_j$ to function exit ($EX$) goes through $bb_i$. 
We say that $bb_i$ dominates $bb_j$ iff \mbox{$bb_i\xrightharpoondown{pre}bb_j \vee bb_i\xrightharpoondown{post}bb_j $}. 
The predominator and postdominator relationships are represented by the trees $T_{pre} $ and $T_{post} $ respectively. 
The dominator graph (DG) is a directed graph that captures all dominance relationships. 
It is obtained by the union of both trees  $DG = T_{pre} \cup T_{post} $, i.e,  by merging edges of both trees.

Given a dominator graph and the fact that a particular $bb$ is covered, this implies that all dominators (predecessors) of $bb$ in DG are also covered.
This allows us to avoid probing basic blocks that do not increase our coverage information.
However, we are interested in moving a step further by leveraging strongly-connected components (SCCs) in the DG.
Each SCC represents a \textit{superblock}, a set of basic blocks with equivalent coverage information.
The superblock dominator graph (SB-DG) is constructed by merging SCCs in the DG.
That is, each node SB in SB-DG represents a SCC in the DG. 
An edge is inserted between $SB_i$ and $SB_j$ iff  $\exists~bb \in SB_i, \exists~bb' \in SB_j$ where $bb$ dominates $bb'$.

Constructing a SB-DG has a number of benefits.
First, it is a convenient tool to measure the coverage information gained from probing any particular basic block.
Second, it enables compressing coverage data by tracking superblocks instead of individual basic blocks.
Finally, it provides flexibility in choosing the best basic block to probe in a superblock. 
We show later in section~\ref{sec:optimized-probe} how this flexibility can be leveraged to reduce instrumentation overhead.

We implemented two instrumentation policies in {\bcov}, namely, \textit{leaf-node} and \textit{any-node}.
We discuss them based on the example depicted in Figure~\ref{fig:probe-pruning}.
In the leaf-node policy, we instrument only the leaves of the SB-DG.
Covering \textit{all} such leaf nodes implies that all nodes in SB-DG are also covered, i.e., achieving 100\% coverage.
However, this coverage percentage is usually infeasible in practice.
Nevertheless, leaf nodes still provide high coverage information which makes the leaf-node policy useful to approximate the coverage of a test suite at a relatively low overhead.

\begin{figure}[t!]
    \centering
    \begin{subfigure}[t]{0.17\textwidth}
        \includegraphics[clip, trim=0.1cm 12.1cm 19.6cm 0.1cm, width=\textwidth]{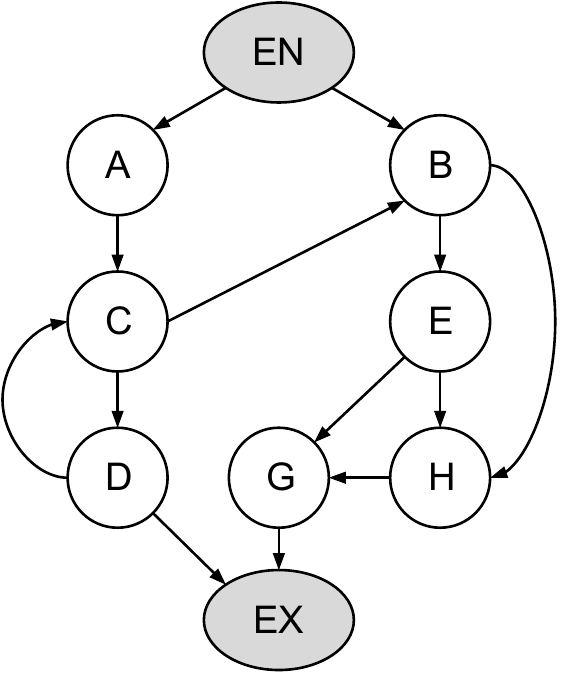}
        \caption{\small CFG}
        \label{fig:3:cfg}
    \end{subfigure}
    \hspace{0.5cm}
    \begin{subfigure}[t]{0.15\textwidth}
        \includegraphics[clip, trim=0.1cm 12.6cm 20.4cm 0.1cm, width=\textwidth]{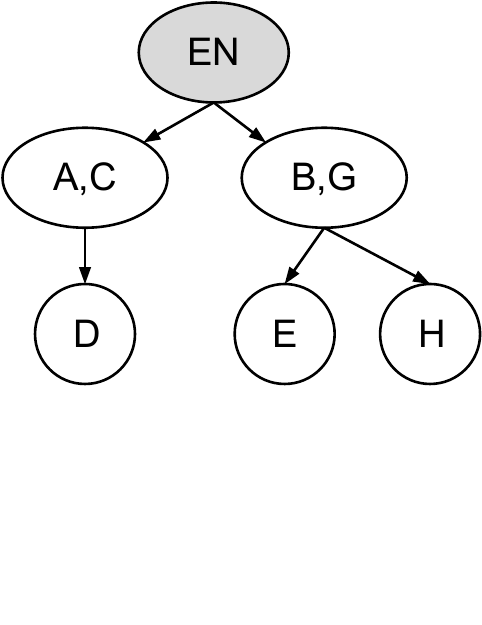}
        
        \caption{\small SB-DG graph}
        \label{fig:3:sbgraph}
    \end{subfigure}
    
    \caption{An example CFG and its corresponding SB-DG.
        First, pre-domominator and post-dominator trees are constructed and merged in a dominator graph (DG). 
        SCCs in DG represent nodes in SB-DG.
        In the \textit{leaf-node} policy, only leaf nodes in SB-DG, namely, D, E, and H, need to be probed. 
        In the \textit{any-node} policy, either A or C need to be additionally probed.
        $EN$ and $EX$ are \textit{virtual} nodes commonly used to simplify dominance analysis.        
    }
    \Description{An example CFG and its corresponding SB-DG}
    \label{fig:probe-pruning}
\end{figure}

Generally, we are also interested in inferring the exact set of covered basic blocks given \textit{any} test input.
This is usually not possible in the leaf-node policy. 
For example, given an input that visits the path $A \rightarrow C \rightarrow B \rightarrow H \rightarrow G$, 
the leaf-node policy can report that the covered set is $\{B,H,G\}$. 
However, this policy can make no statement about the coverage of $A$ and $C$ since they do not dominate the visited probe in $H$.
We address this problem in the any-node policy.
The set of superblocks instrumented in this policy is a superset of those in the leaf-node policy.
More precisely, \mbox{$S_{any}=S_{leaf} \bigcup S_{c}$}.\linebreak
$S_c$ represents the set of \textit{critical} superblocks in the sense that each $sb \in S_c$ can be visited by at least one path in the CFG that does not visit any of its children in the SB-DG.

It is possible to determine $S_c$ using an $\mathcal{O}(|V|+|E|)$ algorithm where $V$ and $E$ are the nodes and edges in the CFG 
respectively.
We refer to \cite{Agrawal1994} for further details.
In Figure~\ref{fig:probe-pruning}, the superblock $\{B,G\}$ is non-critical. However, the superblock $\{A,C\}$ is critical and, consequently, will be probed in the any-node policy.


\section{Control Flow Analysis}
\label{sec:controlflow}
In this section, we first consider the definition of a function at the binary level.
Then, we discuss sliced microexecution, our proposed method for jump table analysis.
\subsection{Function Definitions}
\label{sec:function-definitions}

The notion of function is important to our approach as it determines the scope of CFG and, consequently, the correctness of dominance relationships.
Functions are well-defined constructs in the source code.
However, compiler optimizations such as function splitting and inlining significantly change the layout of corresponding binary-level functions.

Fortunately, these optimizations are not of concern to us as long as \textit{well-formed} function definitions are given to {\bcov}.
A function is defined by the pair $F=(s,z)$ where $s$ and $z$ are start address and byte size respectively.
A function can have a set of \textit{entry} and \textit{exit} points where control flow enters and leaves the function respectively.
We say that a function definition is well-formed if
(1)~its area does not overlap with other functions, and
(2)~all of its basic blocks are reachable only through its entries.

\textbf{Definitions source}.
Our tool uses linker symbols as a source of well-formed function definitions.
These symbols, unlike debug symbols, are available by default in all builds.
In stripped binaries, {\bcov} can read function definitions from call-frame information (CFI) records which can be found in the \texttt{.eh\_frame} section.
This section stores the data necessary for stack unwinding and is part of the loadable image of the binary, i.e., is not stripped.
These records must be available to enable C++ exception handling.
However, they are typically available in C binaries as well since they
are needed for crash reporting, among other tasks.

Note that CFI records might not contain all the functions defined in linker symbols.
For example, developers might exclude CFI records of leaf functions to save memory.
However, we empirically observed that function definitions in CFI records largely match those found in linker symbols.
Additionally, in the unlikely case where CFI records are unavailable,
we may still resort to function identification techniques such as~\cite{Andriesse2017,BYTEWEIGHT2014Bao}.

\textbf{Function entries}.
The main entry of a function is trivially defined by its start address.
Other functions can either \textit{call} or \textit{tail-call} only the main entry.
We have empirically validated this assumption in our dataset.
That is, we have not found any instance where a (direct) function call targets an internal basic block in another function.
However, non-local control transfer mechanisms, such as \texttt{longjmp} and exception handling, violate this assumption.
We refer to possible targets of non-local control transfer as auxiliary function entries.
Such entries are not dominated by, or even unreachable from, the main function entry.
Auxiliary entries of \texttt{longjmp} are identified during CFG construction.
They are simply the successor of each basic block that calls \texttt{setjmp}.

The identification of auxiliary entries used in exception handling is more elaborate.
The Itanium C++ ABI specifies the exception handling standard used in modern Unix-like systems.
Of interest to us in this specification is the \textit{landing pad} which is a code section responsible for catching, or cleaning up after, an exception.
A function can have several landing pads, e.g., it can catch exceptions of different types.
We consider each landing pad to be an auxiliary entry.
Collecting landing pad addresses requires {\bcov} to iterate over all CFI records in the \texttt{.eh\_frame} section.
More specifically, {\bcov} examines all Frame Description Entry (FDE) records looking for a pointer to a language-specific data area (LSDA).
If such a pointer exists, then {\bcov} parses the corresponding LSDA to extract landing pad addresses.

\textbf{Function exits}.
Our tool analyzes the CFG to identify the basic blocks where the control flow leaves a function.
We consider two parameters 
(1)~the type of the control-transfer instruction which can be \texttt{jmp}, \texttt{call}, or \texttt{ret}, and
(2)~whether it is a direct or indirect instruction.
A \texttt{jmp} targeting another function is a tail-call and generally also an exit point.
However, the jump table analysis discussed in section \ref{sec:jumptable} can determine that certain indirect \texttt{jmp} are intra-procedural, i.e., local to the function.
On the other hand,  a \texttt{call} typically returns, i.e, is not an exit point, except for calls to non-return functions.
The non-return analysis implemented in {\bcov} is responsible for identifying such functions.
Finally, we consider all \texttt{ret} instructions to be exit points.

Our model of a function occupying a contiguous code region is simple; yet, we found it to be consistent with our large dataset.
Moreover, it can be augmented with additional analyses to identify function entries and exits.
This provides enough flexibility to handle special situations that might arise in practice, 
for example, using \texttt{ret} to implement indirect calls in Retpoline~\cite{RetpolineTurner}.

\subsection{Jump Table Analysis}
\label{sec:jumptable}

Recovering the targets of indirect control transfer instructions is desirable in several applications such as control-flow integrity.
However, this problem is, in general, undecidable, which means that we can only hope for approximate solutions, i.e. , either to over-approximate or under-approximate the actual set of targets.
Nevertheless, the \texttt{switch} statement in C/C++ remains amenable to precise analysis.
It is commonly implemented as an indirect \texttt{jmp} that is based on single variable indexing into a look-up table. 
This index variable is intra-procedurally bounded.

The analysis of jump tables enables us to
(1)~increase CFG precision,
(2)~instrument jump table data, and
(3)~avoid disassembly errors.
The latter issue is relevant to architectures such as ARM where compilers inline jump table data in the code section.
Fortunately, in x86-64, such data typically reside in a separate read-only section, which enables correct disassembly using linear sweep~\cite{AndriesseUsenixSec16}.

The analysis of jump tables can be challenging as compilers enjoy a lot of flexibility in implementing \texttt{switch} statements.
A jump table can be \textit{control-bounded} by checking the value of the index against a bound condition.
Alternatively, should the expected values be dense, e.g., many values below 16, the compiler might prefer a \textit{data-bounded} jump table, e.g, using a bitwise \texttt{and} with \texttt{0xf}.
Additionally, compilers are free to divide a \texttt{switch} with many \texttt{case} labels into multiple jump tables.
Our goal in this analysis is to recover information about each individual jump table.
This includes its control flow targets and total number of entries.

We propose sliced microexecution, a novel method for jump table analysis which combines classical backward slicing with microexecution~\cite{Godefroid2014}.
The latter refers to the ability to emulate any code fragment without manual inputs.
Basically, for each indirect \texttt{jmp} in a function, {\bcov} attempts to test the sequence of hypotheses depicted in Table~\ref{tab:jump-table-hypothsis}.
If they are invalid then {\bcov} aborts the analysis and considers this \texttt{jmp} to be a tail-call.
Otherwise, {\bcov} proceeds with the actual recovery depending on the type of jump table which can generally either be control-bounded or data-bounded.

\begin{table}[t!]
    \centering
    \small
    \setlength\tabcolsep{3pt}
    \caption{Hypotheses tested, or falsified, to analyze a jump table. Backward slicing answers \#1 to \#3. Microexecution is used to falsify hypotheses and recover the jump table.}
    \label{tab:jump-table-hypothsis}    
    \begin{tabularx}{\columnwidth}{@{}lp{5.2cm}p{2.4cm}@{}}
        \\ 
        &
        \textbf{Hypothesis} &
        \textbf{Action}   \\
        \toprule
        (1)   & Depends on constant base address?    &
        \makecell[tl]{if yes test (2)\\else abort}\\
        (2)   & Is constrained by a bound condition?    &
        \makecell[tl]{if yes test (3)\\else assume (4)} \\
        (3)   & Bound condition dominates jump table?    &
        \makecell[tl]{if yes do recovery\\else assume (4)} \\
        (4)   & \tblentry{Assume jump table is data-bounded}    &
        \makecell[tl]{do recovery and\\try to falsify} \\
        \bottomrule
    \end{tabularx}
\end{table}

We discuss this method based on the example shown in Figure~\ref{fig:jump-table-example}.
First, {\bcov} has to test hypothesis \#1 by backward slicing from \texttt{0x9f719} until it reaches instruction at \texttt{0x9f712} which has a memory dependency.
This dependency has a base address in \texttt{r15}.
So is this base address constant?
Backward slicing for \texttt{r15} shows that it is constant indeed.
Note that a jump table should depend on a single variable used as the index.
The table's base address is a constant determined at compile time.

We move now to test hypothesis~\#2.
It is tested by spawning a \textit{condition slicer} upon encountering each conditional \texttt{jmp}, .e.g, instruction at \texttt{0x9f707}.
This slicer is used to check whether the variable influencing the bound condition is also the jump table index.
This is the case in our example at \texttt{0x9f6f0} where the value in \texttt{r12b} influences both the condition at \texttt{0x9f707} and the jump table index.
Now that a bound condition is found we need to test it against hypothesis~\#3.

A jump table might be preceded by multiple conditional comparisons that depend on the index.
We apply heuristics to quickly discard the ones that can not represent a bound condition, e.g., comparisons with zero.
However, there can still be more than one candidate.
Here, we leverage the fact that a bound condition should dominate the jump table.
Otherwise, a path in CFG would exist where the index value remains unbounded.
We check for dominance during the backward CFG traversal needed for slicing.
Basically, it should not be possible to bypass the bound condition.

Backward slicing produces a program slice (code fragment) which captures the essential instructions affecting the jump table.
This slice represents a univariate block-box function with the index as its input variable.
Modifying the index should trigger behavioral changes especially in the observed jump address at the output.
Assuming that this slice represents a jump table, we reason about its behavior using microexecution.
Also, we try to validate our assumption by widely varying the index.

Before microexecuting a slice, {\bcov} first loads the binary using a built-in ELF loader.
Then, it initializes a valid memory environment for the given program slice.
For example, it allocates memory for the pointer \texttt{[rsp+0x8]} and assigns a valid address to \texttt{rsp}.
It is now possible to start ``fuzzing'' the index.
However, the expected behavior of the slice depends on the type of jump table.

\begin{figure}[t!]
\centering
\begin{lstlisting}[style=custom-x64]
9f6a1: lea    r15,[rip+0xe69e4] ; set table base
                    .
9f6f0: movzx  eax,r12b    ; index is r12b
9f6f4: cmp    r12b,0x5b   ; bound comparison
9f6f8: mov    QWORD PTR [rsp+0x8],rax
|\HiLight|9f6fd: mov    rax,QWORD PTR [rbx]
|\HiLight|9f700: mov    r13,QWORD PTR [rax+0x10]
|\HiLight|9f704: mov    ecx,r13d
9f707: ja     9f880   ; jump to default case
9f70d: mov    rax,QWORD PTR [rsp+0x8]
9f712: movsxd rax,DWORD PTR [r15+rax*4]
9f716: add    rax,r15
9f719: jmp    rax     ; jump to matching case
\end{lstlisting}
\vspace{-2pt}
    \caption[workflow]{Jump table example from \textsf{perl}~v5.28 compiled with \textsf{gcc}~v7.3.
        Highlighted instructions are not part of the backward slice.
        The jump table base is set relatively far at \texttt{0x9f6a1}.}
    \Description{Jump table example from perl v5.28 compiled with gcc v7.3}
    \label{fig:jump-table-example}
\end{figure}

In control-bounded jump tables, a change in behavior must be observed in the intervals $[0,b)$ and $(b,\plusinfty)$ where $b$ is the bound constant.
This constant is located in the first instruction that sets the flags before the bound condition.
In our example, this is the instruction at \texttt{0x9f6f4}.
{\bcov} tests 24 index values in total, 8 of which are sampled from $[0,b]$ including $0$, $b-1$, and $b$.
The remaining 16 values increase exponentially, in powers of 2, starting from $b+1$.
We found this scheme to give us high confidence in the results.

The jump table is expected to target an instruction inside the current function for most inputs in $[0,b)$.
On the other hand, the jump table should not be reachable for all inputs in $(b,\plusinfty)$.
That is, the bound condition should redirect control flow to the default case.
Should the behavior of the program slice not match what we expect from a control-bounded jump table,
then we abort and assume that it is data bounded.
Note that we are not strict about the behavior for input $b$ since the bound condition might check for equality.

Assuming that a given indirect \texttt{jmp} represents a data-bounded jump table, we need
effective techniques to (1) stop backward slicing, (2) validate our assumption, and (3) explore
the bound limits.
Note that compilers might use more than one bitwise instruction to bound the index. 
Moreover, developers might prefer computed gotos over switch statements. ~\footnote{Computed gotos is a \textsf{gcc} extension to C which is also supported in \textsf{clang}.}
In this case, they need to assume responsibility for checking index bounds. 

To cope with this implementation diversity, bcov continues backward slicing as long as the current slice depends on only one variable. 
For example, assume that \texttt{rax} holds the index and is later used as a base register to read from memory. 
This means that the current slice would depend on \texttt{rax} as well as the variable accessed in memory. 
Backward slicing would stop before this increase in dependencies.
Then, \bcov{} executes the program slice 24 times, each time increasing the index exponentially while setting the least significant bits to one.
This allows us to explore the bound limits in the common case of bitwise \texttt{and} with a bitmask like \texttt{0xf}.
Other bit patterns are also tried to better penetrate combinations of bitwise instructions.
Our key insight is that we should not have full control over the jump target.
That is, arbitrary change in the index should be reflected in a \textit{constrained} change in the jump target.
Additionally, jump targets need to be located in the current function similar to the case of control-bounded jump tables.
Should the program slice withstand these diverse tests, then we can be highly confident that it represents a jump table.

Our evaluation shows that sliced microexecution is precise and robust against various compiler optimizations.
It allowed {\bcov} to reliably recover the jump tables in the core loop of the Python interpreter, located in function \texttt{\_PyEval\_EvalFrameDefault}. Note that these jump tables are compiled from complex computed gotos.


\section{Static Instrumentation}
\label{sec:instrumentation}
In this section, we first consider a strategy to reduce instrumentation overhead by carefully selecting a basic block to probe in a superblock.
Then, we discuss handling short basic blocks by means of hosting their detours in larger neighboring basic blocks.

\subsection{Optimized Probe Selection}
\label{sec:optimized-probe}

Generally, probing a BB requires inserting a detour targeting its designated trampoline.
A detour occupies 5 bytes and can either be a direct \texttt{jmp} or \texttt{call}.
Consequently, one or more original instructions must be relocated to the trampoline.
This \textit{relocation} overhead varies due to the instruction-size variability in x86-64.
Note that a pc-relative \texttt{mov}, which occupies 7 bytes, represents an unavoidable
overhead for updating coverage data in each trampoline.
Hence, our goal is to reduce the relocation overhead.

To this end, we iterate over all BBs in a superblock and select the one expected to incur the lowest overhead.
First, we have to establish whether a detour can be accommodated in the first place.
A BB that satisfies $s + p < 5 $ is considered a \textsf{guest}, where $s$ and $p$ are the byte size and padding size respectively.
A superblock that contains only \textsf{guest} BBs is handled via detour hosting (\S\ref{sec:detour-hosting}).
Now we examine the type and size of the last instruction of each BB and whether the BB is targeted by a jump table.
These parameters are translated to the types depicted in Table~\ref{tab:probe-overhead}.
These BB types are organized in a total order.
This means, for example, we strictly prefer a \textsf{long-call} over a \textsf{long-cond} should both exist in the same superblock.
This type order is primarily derived from empirical observation.
However, we did not necessarily experiment with all possible combinations.
Preferring \textsf{long-call} over \textsf{short-call} should be intuitive.
The latter incurs an additional overhead for relocating at least one instruction preceding the \texttt{call}.

\begin{table}[t!]
    \centering
    \renewcommand{\arraystretch}{1.2}
    \setlength\tabcolsep{4pt}
    \small
    \caption{BB classification used in probe selection. Types are shown in ascending order based on expected relocation overhead. 
    The terms \textit{long} and \textit{short }are relative to detour size (5 bytes).
    Short types require relocating preceeding (RP) instruction(s).
    } 
    \label{tab:probe-overhead}    
    \begin{tabularx}{\columnwidth}{@{}lll@{}}
        \\
        \textbf{Type} &
        \textbf{RP} &
        \textbf{Relocation overhead} \\
        
        \toprule
        
        \textsf{return} & maybe & Can be only 1 byte depending on the padding \\
        \textsf{long-jump} & no & Size of \texttt{jmp} instruction which is $ \ge $ 5 bytes \\
        \textsf{long-call} & no & Size of \texttt{call} instruction which is $ \ge $ 5 bytes  \\
        \textsf{jump-tab} & no & Size of \texttt{jmp} instruction to original code (5 bytes)  \\
        \textsf{short-call} & yes & Similar to \textsf{long-call} but with RP overhead added \\
        \textsf{short-jump} & yes &Similar to \textsf{long-jump} but with RP overhead added\\
        \textsf{internal} & maybe & Size of relocated instruction(s) inside the BB \\
        \textsf{long-cond} & no & Rewriting incurs a fixed 11 byte overhead\\
        \textsf{short-cond} & yes & Similar to \textsf{long-cond} but with RP overhead added\\
        \bottomrule
    \end{tabularx}
\end{table}

We observed that \textsf{return} basic blocks are usually padded (55\% on average).
Padding size is often more than 3 bytes which translates to a relocation overhead of only one byte - the size of a  \texttt{ret} instruction.
Also, favoring \textsf{long-jmp} over \textsf{long-call} provided around 3\% improvement in both relocation and performance overheads.
On the other hand, \textsf{short-call} had only a slight advantage over \textsf{short-jmp}.
This might be due to the fixed 2-byte size of the latter, which leads to relocating more instructions.
However, our experiments were not always conclusive, e.g., between \textsf{jump-tab} and \textsf{short-call}.

Relocating an instruction depends on its relation to the PC (called \textsf{rip} in x86-64).
Position-independent instructions can simply be copied to the trampoline.
However, we had to develop a custom rewriter for position-dependent instructions.
The rewriter preserves the exact semantics of the original instruction whether it explicitly or implicitly depends on \texttt{rip}. 
For example, a \textsf{long-cond} instruction will be rewritten in the trampoline to a matching sequence consisting of a \textsf{long-cond} (6 bytes) and a \textsf{jmp} (5 bytes).

Jump table instrumentation has the unique property of preserving the original code.
It is a data-only mechanism that enables us to probe even one-byte BBs.
However, for it to be applicable, a BB has to be targeted by a \textit{patchable} jump table.
A jump table is patchable if its entries are either 32-bit offsets or absolute addresses.
We observed that about 92\% of more than 46,000 jump tables in our dataset are patchable.
In fact, we found that 8-bit and 16-bit offsets are only used in \texttt{libopencv\_core}.

Finally, our probe selection strategy is effective in reducing relocation overhead.
However, it is not necessarily optimal.
We observed high variance in the padding of BBs of type \textsf{return}, i.e., \textsf{return} is not always the best choice.
Also, instrumenting a loop head can unnecessarily trigger multiple coverage updates. A loop-aware strategy might reduce performance overhead by choosing a BB outside the loop as an alternative. Such optimizations are left for future work.

\subsection{Detour Hosting}
\label{sec:detour-hosting}

The instruction-size variability in x86-64 suggests that some BBs are simply too short to safely insert a detour without overwriting other BBs.
In our dataset, we found that about 7\% of all BBs are short (size < 5 bytes).
Left without a probe, we risk losing coverage information of a particular short BB and, potentially, all of its dominators.
One possible solution is to relocate the entire function to a larger memory area.
However, this is costly in terms of code size and the engineering effort required to fix relocated code references.
For example, throwing an exception from a relocated function without fixing its corresponding CFI record might lead to abrupt process termination.

The method adopted in {\bcov} is \textit{detour hosting}.
It significantly reduces the relocation overhead while preserving the stability of code references at basic-block level.
Here, the size of a guest BB needs to be at least 2 bytes, which is enough to insert a short detour targeting a reachable host BB, i.e., within about $\pm$128 bytes.
The host BB must be large enough to accommodate two regular detours, i.e., at least 10 bytes.
The first detour targets the host trampoline while the other detours would target the trampolines of their respective guests.
Note that we can safely overwrite padding bytes of both the guest and host.
Also, the host does not need to be entirely relocated.
Relocating a subset of its instructions might be sufficient.

Figure~\ref{fig:detour-hosting} depicts a detour hosting example.
It involves a guest BB consisting of a single indirect \textsf{call} (3 bytes).
The tricky part about a \textsf{call} is that its return address must be preserved.
A \texttt{sub} instruction (5 bytes) is used to adjust the return address in the trampoline from \texttt{0xad67fd} to its original value of \texttt{0xad6803}.
This adjustment also clobbers the CPU flags which is safe since they are not preserved across function calls in the x86-64 ABI.
Note that this is the only case where we modify the CPU state.

Now we have the following allocation problem: given a guest $g$ and a set of suitable hosts $H=\{h_1,h_2,..,h_n\}$, find the host $h_i$ whose selection incurs minimal overhead.
Also, we are interested in the more general formulation: given a set of guests $G=\{g_1,g_2,..,g_k\}$ and a set of hosts $H=\{h_1,h_2,..,h_n\}$, where each host is suitable for at least one guest, find a function mapping $M:G\rightarrow H$ such that the overhead is minimal.
We approach this problem using a greedy strategy where we prefer, in this order,
(1)~packing more guests in a single host,
(2)~a host already selected for instrumentation over an otherwise intact host,
(3)~a host that is closer to the guest.
Basically, for each guest, we iterate over all reachable BBs.
A BB can offer a hosting offset, if possible.
A higher offset means that more guests are packed in this host.
The initial offset is 5 bytes from the start of the host.
Should the offered offsets be equal, we look into (2) to avoid, as much as possible, relocating otherwise intact BBs.
Finally, should both (1) and (2) be equal, then we look into (3) to improve the code cache locality.

It is not possible to probe one-byte guests.
Also, a suitable host might not be found.
However, we can still reduce the loss in coverage information in such cases.
To this end, we try to probe all of the immediate predecessors of the current SB, containing the guest, in the SB-DG.
However, this does not necessarily entail adding more probes.
For example, additional probes are unnecessary in the leaf-node policy should the current SB have siblings in the SB-DG.
In the any-node policy, on the other hand, the predecessors might be probed already.

Our detour hosting strategy targets a sweet spot balancing performance and relocation overheads.
It achieves a hosting ratio of 1.2 guests per host on average.
Also, it was able of hosting up to 14 guests in a single host.
Around 80\% of the hosts are already probed. 
That is, relocation overhead is expected for them anyway.
Finally, it allowed {\bcov} to host about 94\% of all the guests.

\begin{figure}[t!]
	\centering
	\vspace{-4pt}
	\begin{subfigure}[t]{\columnwidth}
\begin{lstlisting}[style=custom-x64]
ad67f3: jmp   ad6803
ad67f5: nop   [multi-byte]
ad6800: call  QWORD PTR [rax+0x58]
\end{lstlisting}
		\vspace{-4pt}
		\caption{original code}
		\label{fig:5:original}
	\end{subfigure}
	\begin{subfigure}[t]{\columnwidth}
\begin{lstlisting}[style=custom-x64]
ad67f3: jmp   1d31afa  ; jump to relocated host
ad67f8: call  1d31b39  ; hosted detour
ad67fd: nop   DOWRD PTR [rax]
ad6800: jmp   ad67f8   ; jump to hosted detour
\end{lstlisting}
		\vspace{-4pt}
		\caption{patched code}
		\label{fig:5:patch}
	\end{subfigure}
	\hfill
	\begin{subfigure}[t]{\columnwidth}
		\vspace{-4pt}
\begin{lstlisting}[style=custom-x64]
1d31b39: mov  BYTE PTR [rip+0x4f8d01],1
1d31b40: sub  QWORD PTR [rsp], -6
1d31b45: jmp  QWORD PTR [rax + 0x58]
\end{lstlisting}
		\vspace{-4pt}
		\caption{trampoline}
		\label{fig:5:guest-trampoline}
	\end{subfigure}
	
	\caption{Detour hosting example taken from \textsf{llc} v8.0 compiled with \textsf{clang} v5.0.
		\textbf{(a)} host is a short \texttt{jmp} at \texttt{0xad67f3} followed by 11 padding bytes.
		\textbf{(b)} inserting 2 detours leaves 3 padding bytes.
		\textbf{(c)} return address adjusted at \texttt{0x1d31b40}. Original \texttt{call} at \texttt{0xad6800} is rewritten to a matching \textsf{jmp} at \texttt{0x1d31b45}}
    \Description{Detour hosting example taken from llc v8.0 compiled with clang v5.0.}
	\label{fig:detour-hosting}
\end{figure}


\section{Implementation}

We implemented our approach in the tool {\bcov}.
Our tool accepts an ELF module (executable or shared library) as input.
It starts with a set of module-level analyses such as reading function definitions, parsing CFI records, and building the call graph.
Our non-return analysis implementation is similar to~\cite{Meng:ISSTA2016}.
We omit the details as they are not part of our core contribution.

Then, {\bcov} moves to function-level analyses such as building the CFG (including jump tables), dominator trees, and superblock dominator graph.
Probes are determined based on the instrumentation policy set by the user.
{\bcov} can be used for patching or coverage reporting.
The latter mode requires a data file dumped from a patched module.
The instrumentation policy used for coverage reporting must match the one used for patching. 

We implemented the modern SEMI-NCA dominator tree algorithm~\cite{Georgiadis2005} and Tarjan's classical SCC algorithm.
We used \textsf{capstone}~\cite{CapstoneEngine} for disassembly and implemented a wrapper around \textsf{unicorn}~\cite{UnicornEngine} for microexecution.
In total, this required about 17,000 LoC in C++ (testing code excluded).
The run-time library \textsf{bcov-rt} is implemented in C in $\sim$250 LoC.

\section{Evaluation}
\label{sec:evaluation}

Our evaluation is guided by the following research questions:

\begin{description}
    \itemsep-.1em
    \item[RQ1] Can {\bcov} transparently scale to large real-world binaries?
    \item[RQ2] What is the instrumentation overhead in terms of performance, memory, and file size?        
    \item[RQ3] Have we pushed the state of the art in jump table analysis?    
    \item[RQ4] To what extent can {\bcov} provide better efficiency in comparison to its direct alternatives, namely, DBI tools?    
    \item[RQ5] Can {\bcov} accurately report binary-level coverage?    

\end{description}

For evaluation, we selected eight modules from popular open-source packages offering diverse functionality.
They are summarized in Table~\ref{tab:selected-benchmarks}.
We compiled each module using four compilers in three different build types.
Specifically, we used the compilers \textsf{gcc-5.5}, \textsf{gcc-7.4}, \textsf{clang-5.0}, and \textsf{clang-8.0}.
This gives us a representative snapshot of the past three years of developments in \textsf{gcc} and \textsf{clang} respectively.
The build types are \textsf{debug}, \textsf{release}, and \textsf{lto}.
The latter refers to link-time optimizations.
Compiler optimizations were disabled in \textsf{debug} builds and enabled in \textsf{release} and \textsf{lto} builds.
Enabled optimizations depend on the default options of their respective package which can  be at levels \texttt{O2} or \texttt{O3}.

This results in 12 versions of each module and a total of 95 binaries.
\footnote{Compiling \textsf{llc} with \textsf{gcc-5.5} in \textsf{lto} build resulted in a compiler crash.}
Our tool was capable of patching 88 binaries without modifying the build system.
However, we had to modify the linker script in 7 instances where relocating ELF segment headers was not possible. 
We instructed the linker to leave 112 bytes, which is enough for our segment headers, after the original segment headers.
This change is small affecting only one line in the linker script.
The \mbox{\textsf{bcov-rt}} runtime was injected using the LD\_PRELOAD mechanism.
All experiments were conducted on an Ubuntu 16.04 PC with Intel\textsuperscript{\textregistered} i7-6700 CPU and 32GB of RAM.

\subsection*{RQ1. Scalability and Transparency}
Our choice of subjects directly supports our claim regarding scalability.
Figure~\ref{fig:code-size-comp} shows a comparison in terms of code size relative to \textsf{objdump}, a commonly used subject in binary analysis research.
Code size is measured using the popular \texttt{size} utility.
Note that {\bcov} can analyze and patch our largest subject, \textsf{llc}, in $\sim$30 seconds.
In our experiments, we used {\bcov} to instrument all functions  available in the \texttt{.text} section.
More than $1.6\times10^6$ functions have been instrumented across 95 binaries.
The policies leaf-node and any-node have been applied separately, i.e., subjects were instrumented twice.

\begin{table}[t!]
	\centering
	\setlength\tabcolsep{4pt}
	\small
	\caption{Selected evaluation subjects. Used recent package versions.}
    \label{tab:selected-benchmarks}    
	\begin{tabularx}{\columnwidth}{@{}llll@{}}
        \\
		\textbf{Module} &
		\textbf{Package} &
		\textbf{Lang.}  &
		\textbf{Domain} \\		
		\toprule
		
		\textsf{gas} & binutils-2.32 & C & Assemblers\\
		\textsf{perl} & perl-5.28.1 & C & Interpreters \\
		\textsf{python} & cpython-3.7.3 & C & Interpreters\\
		\textsf{libMagickCore} & ImageMagick-7.0.8 & C & Image processing\\
		\textsf{ffmpeg} & FFmpeg-4.1.3 & C & Video processing\\
		\textsf{libxerces-c-3.2} & xerces-c-3.2.2 & C++ & XML processing \\
		\textsf{libopencv\_core} & opencv-4.0.1 & C++ & Computer vision\\
		\textsf{llc} & llvm-8.0.0 & C++ & Compilers\\
		\bottomrule
	\end{tabularx}
\end{table}

Transparency is important in coverage instrumentation.
This practically means that {\bcov} should not introduce test regressions.
We evaluated this criterion by replacing original binaries with instrumented versions and re-running their test suites.
\textit{Our instrumentation did not introduce any regressions} despite the fact that (1) we systematically patch all functions, even compiler-generated ones, and (2)
our benchmark packages include extensive test suites.
For example, the \textsf{perl} test suite runs over one million checks.

\begin{figure}[t!]
	\centering
	\includegraphics[clip, trim=0.47cm 0.4cm 0.45cm 0.65cm, width=0.8\columnwidth]{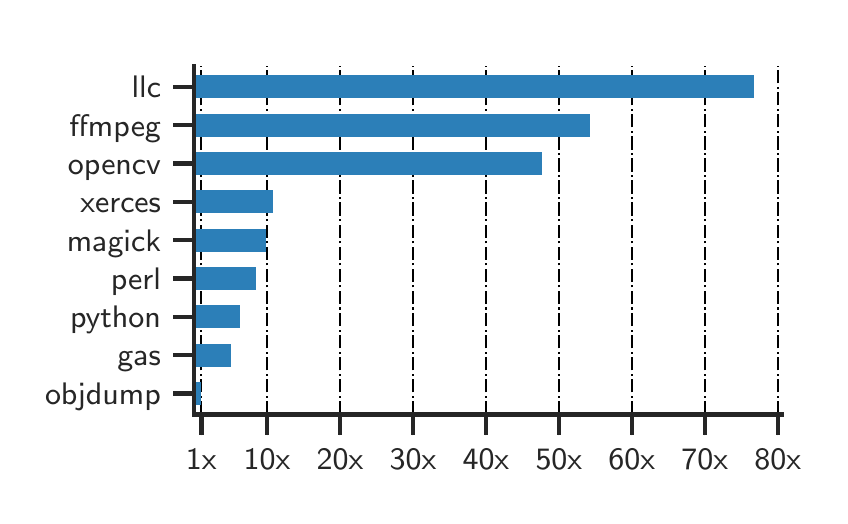}
	\caption{Comparing the code size of our subjects to \textsf{objdump} (code size about 339KB). Code size reported with GNU \textsf{size} utility.}
	\Description{Comparing the code size of our subjects to the size of objdump}
	\label{fig:code-size-comp}
\end{figure}

\subsection*{RQ2. Instrumentation Overhead}

\begin{figure*}[t!]
	\centering
	\small
	\begin{subfigure}[t]{0.29\textwidth}
		\includegraphics[clip, trim=0.48cm 0.4cm 1.4cm 0.65cm, width=\textwidth]{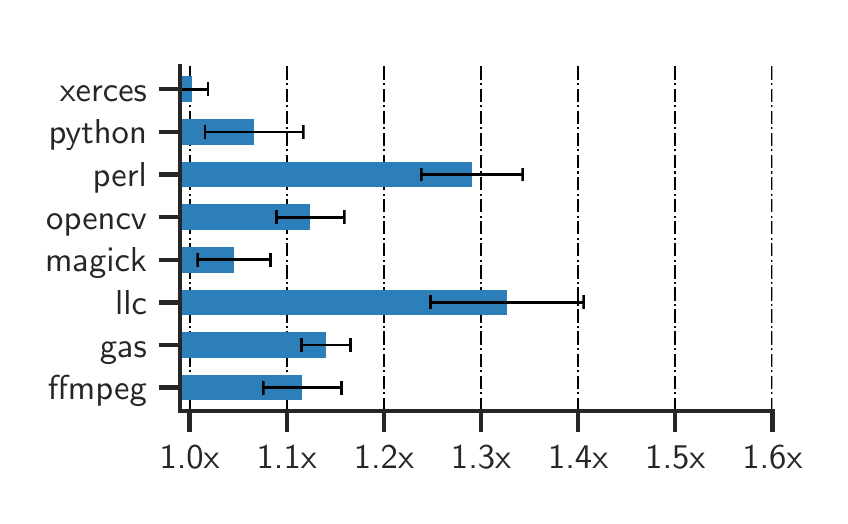}
		\caption{performance overhead}
		\label{fig:performance-overhead}
	\end{subfigure}
	\hspace{10pt}
	\begin{subfigure}[t]{0.29\textwidth}
		\includegraphics[clip, trim=0.47cm 0.42cm 1.65cm 0.65cm, width=\textwidth]{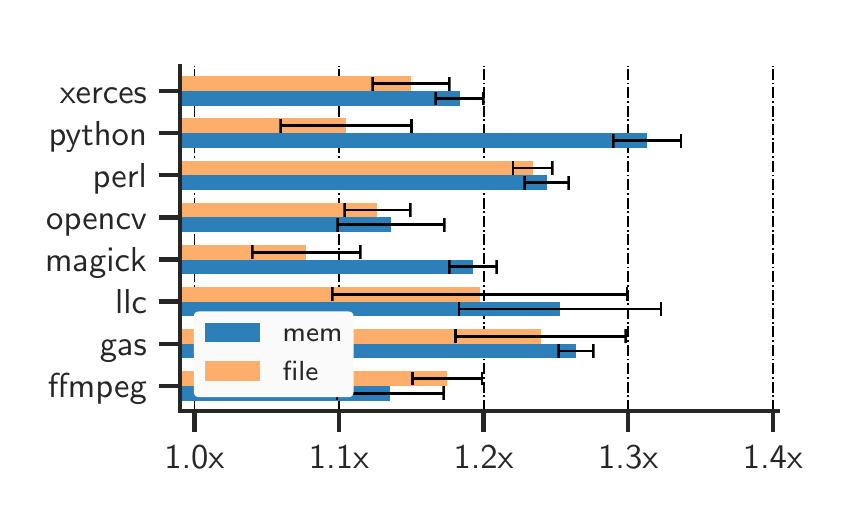}
		\caption{size overhead}
		\label{fig:size-overhead}
	\end{subfigure}
	\hspace{10pt}
	\begin{subfigure}[t]{0.3\textwidth}
		\includegraphics[clip, trim=0.48cm 0.36cm 0.75cm 0.65cm, width=\textwidth]{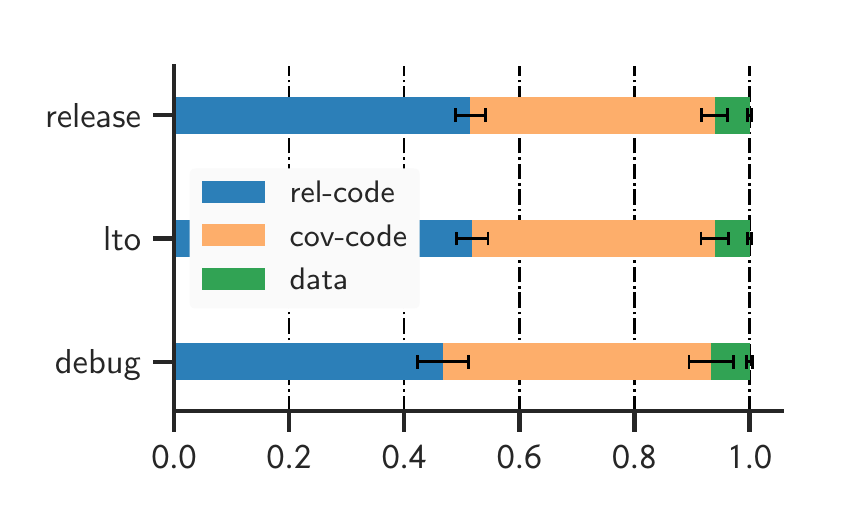}
		\caption{overhead distribution}
		\label{fig:overhead-dist}
	\end{subfigure}
	
	\caption{Overhead of the any-node instrumentation policy. (a) performance overhead accounts for instrumentation and dumping coverage data, (b) memory and file size overhead (c) distribution of memory overhead between code (relocated  and coverage update) and coverage data. }
	\Description{Overhead of the any-node instrumentation policy}
	\label{fig:overhead-results}
\end{figure*}

\begin{figure*}[t!]
	\centering
	\small
	\begin{subfigure}[t]{0.27\textwidth}
		\includegraphics[clip, trim=0.48cm 0.4cm 1.71cm 0.65cm, width=\textwidth]{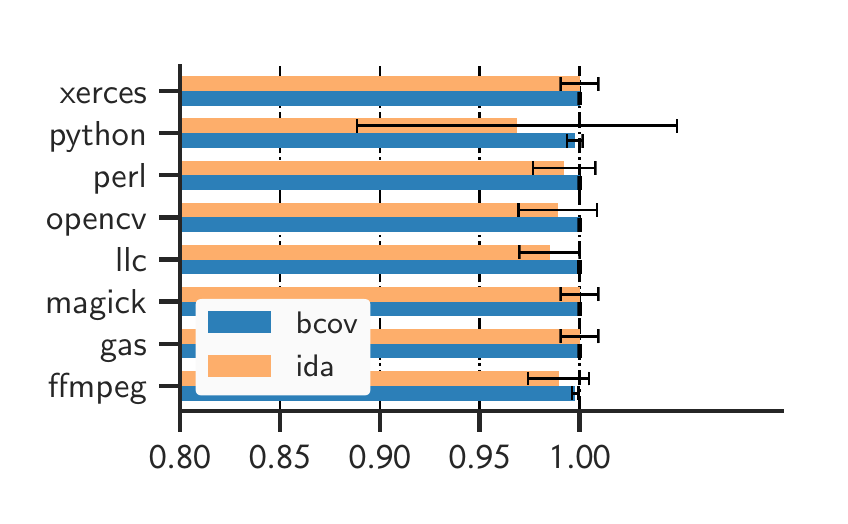}
		\caption{\textsf{clang} binaries}
		\label{fig:jumptab-clang}
	\end{subfigure}
	\hspace{10pt}
	\begin{subfigure}[t]{0.28\textwidth}
		\includegraphics[clip, trim=0.48cm 0.4cm 1.6cm 0.65cm, width=\textwidth]{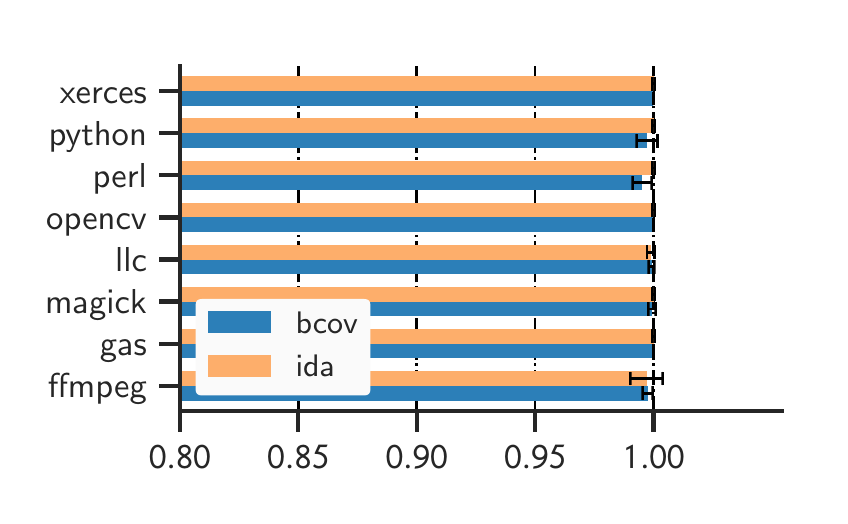}
		\caption{\textsf{gcc} binaries}
		\label{fig:jumptab-gcc}
	\end{subfigure}
	\hspace{10pt}
	\begin{subfigure}[t]{0.28\textwidth}
		\includegraphics[clip, trim=0.48cm 0.38cm 1.6cm 0.8cm, width=\textwidth]{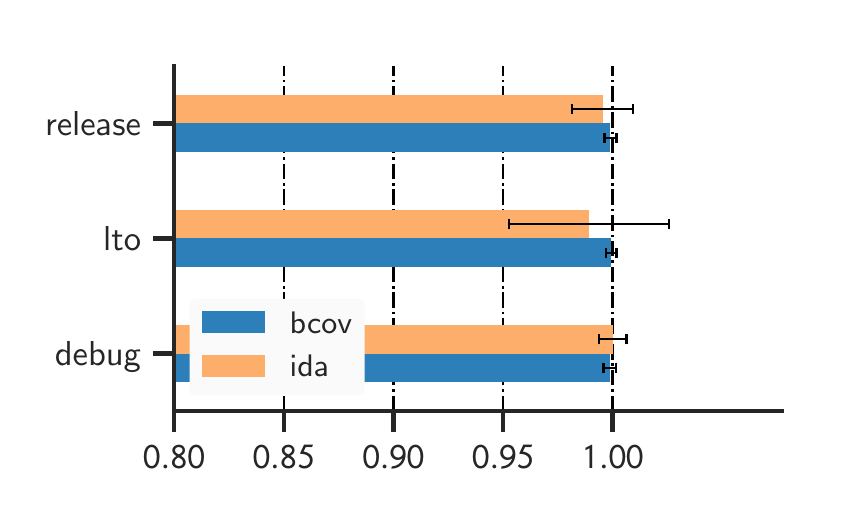}
		\caption{build type}
		\label{fig:jumptab-build-type}
	\end{subfigure}
	
	\caption{Normalized jump table analysis results in comparison to IDA Pro. \textbf{(a)} IDA Pro shows significant  variance on \textsf{clang} binaries. \textbf{(b)} both tools are comparable on \textsf{gcc} binaries. \textbf{(c)} varying the build type did not affect {\bcov}.}
	\Description{Normalized jump table analysis results in comparison to IDA Pro}
	\label{fig:jumptab-results}
\end{figure*}

Figure~\ref{fig:overhead-results} depicts the instrumentation overhead of the any-node policy relative to the original binaries. 
We omit the detailed evaluation of the leaf-node policy because of the lack of space.
The average performance overheads of the leaf-node and any-node polices are 8\% and 14\% respectively.
The overhead is measured based on the wall-clock time required to run individual test suites, .e.g, run ``\textsf{make test}'' to completion.
This covers the overhead associated with instrumentation and dumping coverage data to disk.
The latter overhead varies depending on the number of processes spawned during testing.

For example, all \textsf{opencv} tests are executed within a single process that dumps coverage data only once.
On the other hand, unit testing of \textsf{llc} spawns over 7,500 processes in about 40s.
This results in dumping $\sim$4\,GB of coverage data which significantly contributes to the overall delay.
Online merging of coverage data might reduce this disk IO overhead.
To give a better intuition, we note that without online merging, \textsf{llvm-cov} would dump over 320GB of coverage (and profiling) data for the same benchmark. 

To gain a better insight into the distribution of performance overhead, we experimented with the following variants of the any-node policy:

\begin{itemize}
	\item Detour only (DO). Here, we enabled only inserting detours and relocating overwritten instructions to the trampolines.
	\item Detour hosting (DH). Similar to DO, but we also enabled detour hosting for short basic blocks.
	\item Coverage update (CU). Similar to DH, but we also enabled inserting coverage update code in the trampoline. 
\end{itemize}

Note that the last variant is equivalent to the any-node policy.
We separately used each variant to patch all subjects in 4 different release builds.
Then, we ran their respective test suites to measure the performance overhead.
Interestingly, DO alone accounts for about 96\% of the overhead on average.
That is, DH and CU contributed only marginally.
This result suggests that inlining coverage update code, and thereby eliminating the need for trampolines,  may lead to substantial performance savings.

The average memory and file size overheads introduced by {\bcov} are 22\% and 16\% respectively.
We measure the memory overhead relative to the size of loadable ELF segments only. 
Recall that {\bcov} does not affect the run-time heap or stack.
The fact that other static instrumentation techniques need to duplicate the code segment~\cite{Anand:EuroSys2013,Laurenzano2010}, suggests that the overhead of {\bcov} is reasonable.
Coverage data represents only 6\% of the memory overhead.
It is worth noting that compiler optimizations can force {\bcov} to relocate more instructions.
This might be due to emitting smaller basic blocks.
However, our static instrumentation techniques are effective in reducing the difference in relocation overhead between debug and optimized builds as shown in Figure~\ref{fig:overhead-dist}.

\subsection*{RQ3. Jump Table Analysis}
Evaluating sliced microexecution requires comparing {\bcov} with representative binary analysis tools.
However, it was not possible to compare with BAP~\cite{BAP:CAV2011Brumley} and \textsf{angr}~\cite{angr:Shoshitaishvili2016} which are the leading academic tools.
BAP does not have built-in support for jump table analysis, while \textsf{angr} (v8.18.10) crashed on \textsf{opencv} and \textsf{llc} binaries.
For the remaining binaries, \texttt{angr} reported significantly fewer jump tables compared to IDA Pro.
Therefore, we compare {\bcov} only with IDA Pro (v7.2).
This should not affect our results since IDA~Pro is the leading industry disassembler.

Next, we have to establish the ground truth of jump table addresses --- specifically, the addresses of their indirect \texttt{jmp} instructions.
This is challenging as compilers do not directly emit such information.
Therefore, we conducted a differential comparison.
We observed that {\bcov} and IDA Pro agree on the majority of jump tables including their targets, so we manually examined the remaining cases where they disagree.
Both tools did not report false positives.
That is, they only missed jump tables.
This is expected in {\bcov} as repeated microexecution inspires high confidence in its results.
Therefore, our ground truth is the union of jump table addresses recovered by both tools.

Figure~\ref{fig:jumptab-results} depicts the recovery percentages relative to this ground truth consisting of over 46,000 jump tables.
We control for different factors affecting compilation.
We observed that IDA Pro delivers lower accuracy on \textsf{clang} binaries compared to \textsf{gcc} binaries, and its accuracy was affected by compiler optimizations.
In comparison, {\bcov} demonstrates higher robustness across the board.

\subsection*{RQ4. Comparison with DBI Tools}

\begin{table*}[t!]
	\centering
	\setlength\tabcolsep{8.5pt}
	\small
	\caption{Evaluating the accuracy of \textsf{bcov} based on \textsf{drcov} traces. 
    We show the number of processes spawned during testing, corresponding dump sizes in MB,
    and the total number of BBs and their instructions in original binaries.
    Both tools dump one coverage file per process.
    For each subject, we list the average/maximum of true positives (TP).
    FPs and FNs are also considered by listing the average precision and recall respectively.		 
    DR could not complete the test suite runs of \textsf{perl} and \textsf{python}.
    Omitted \textsf{opencv} as \textsf{drcov}'s data was invalid due to a bug. 
    }
    \label{tab:coverage-accuracy}        
	\begin{tabularx}{0.99\textwidth}{@{}lllllllllll@{}}
        \\
		\textbf{Module} &
		\textbf{process \#} &
		\textbf{drcov size} &
		\textbf{bcov size} &
		\textbf{BB} &
		\textbf{Inst.} &
		\textbf{TP BB} &
		\textbf{TP Inst.} &
		\textbf{Precision} &
		\textbf{Recall} \\		
		\toprule
		
		xerces   & 80     & 12.34     & 4.32     & 116378   & 420096   & 9523.2~/~21927    & 40651.2~/~92144     & 99,98\%   &  99.42\%   \\ 
		magick   & 58     & 7.71      & 2.90     & 125521   & 521107   & 5689.4~/~20709    & 21614.9~/~83444     & 99,98\%   &  99.94\%    \\ 
		llc      & 7862   & 3481.97   & 4176.16  & 1067151  & 4343021  & 45184.5~/~90952   & 257209.5~/~461656   & 99,98\%   &  99.68\%  \\ 		
		gas      & 1235   & 71.94     & 38.56    & 60511    & 220447   & 2916.4~/~5015     & 11045.8~/~19578     & 99.93\%   &  99.67\%   \\ 
		ffmpeg   & 3309   & 423.45    & 762.39   & 496404   & 3050228  & 9682.0~/~14489    & 41439.1~/~63591     & 99.98\%   &  99.94\%   \\ 
		
		\bottomrule
	\end{tabularx}
\end{table*}

Pin and DynamoRIO (DR) are the most popular DBI tools.
Both act as a process virtual machine which instruments programs while JIT-emitting instructions to a code cache.
This complex process creates the following sources of overhead: 
(1) JIT optimization, and (2) client instrumentation.
To evaluate this overhead on our test suites, we installed the latest stable releases of both tools, namely, Pin v3.11 and DR v7.1.
We then replaced each of our subjects with a wrapper executable.
In the case of shared libraries, we replaced their test harness with our wrapper.
The test system would now run our wrapper, which in turn runs its corresponding original binary but under the control of a DBI tool. 
The wrapper reads a designated environment variable to choose between Pin and DR.

\cref{fig:dbi-tools-comparison} depicts the performance overhead of Pin and DR without client instrumentation.
It also shows the overhead of DR after enabling \textsf{drcov}, its code coverage client.
Note that Pin does not have a similar coverage client built-in.
The overhead is measured relative to original binaries and is averaged for four different release builds.
Both tools introduced regressions on \textsf{perl} and \textsf{python}.
However, DR made tests hung on \textsf{perl} and crashed on the \textsf{python} test suite. 
This highlights the challenges of maintaining transparency in DBI tools. 
Note that the DBI overhead of executable subjects is significantly higher than that of shared libraries. 
This can be attributed to the start-up delay which dominates in short-running tests.
The average performance overheads of Pin and DR are 29.1x and 4.1x respectively.
Enabling \textsf{drcov} increases the average overhead to 7.3x.
For the same benchmarks, {\bcov} introduced an average overhead of 11\% only.
Our experiments show that {\bcov} provides significantly better performance, transparency, and usability.

\begin{figure}[t!]
    \centering
    \includegraphics[clip, trim=0.48cm 0.4cm 0.49cm 0.73cm, width=0.9\columnwidth]{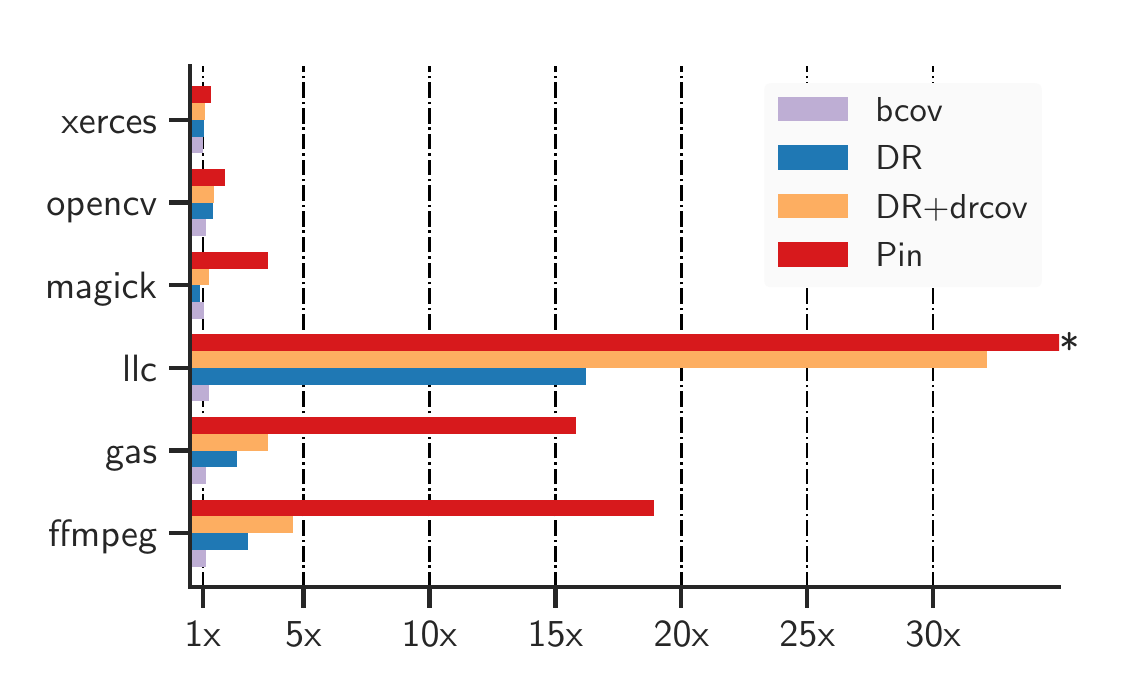}
    \caption{Comparing the performance overhead of Pin, DynamoRIO (DR), \textsf{drcov}, and {\bcov}. Omitted \textsf{perl} and \textsf{python} as DR was  unable of completing their test suite runs. 
        (*) The actual overhead of Pin for \textsf{llc} is over 130x.}
    \Description{Comparing the performance overhead of Pin, DynamoRIO (DR)}
    \label{fig:dbi-tools-comparison}
\end{figure}

\subsection*{RQ5. Coverage Report Accuracy}
We evaluate the accuracy of the reported coverage by tracing binaries that are instrumented by {\bcov}. 
We use the any-node policy because of its precision.
Note that comparing the coverage of original binaries separately to instrumented ones will likely introduce errors that are only caused by non-determinism.
For example, repeatedly printing a simple: ``Hello World'' using \textsf{perl} will produce different instruction traces.

Initially, we obtained the ground truth traces using Intel~PT (IPT).
To this end, we collected about 2,000 sample tests from our test suites.
Running these tests produces 104GB of IPT data and 444MB of {\bcov} coverage data.
We used the standard \textsf{perf} tracing facilities in kernel v4.15 and later kernel v5.3.
We tried many IPT configurations and restricted ourselves to tests terminating in $ \le $ 5 seconds.
Despite these efforts, we could not reliably evaluate {\bcov} due to non-deterministic loss in IPT data. 
After all, disks might just be incapable of keeping up with the CPU~\cite{IntelPTLinuxDocs}.

We then turned to \textsf{drcov} to obtain the ground truth.
This DR client dumps the address of encountered basic blocks (BB) heads, i.e., first instruction.
We leverage the fact that our instrumentation does not modify BB heads.
Based on this, we expect BBs reported as covered by {\bcov} to appear in \textsf{drcov}'s trace.
We consider these BBs to be true positives (TP).
On the other hand, a BB reported by bcov that was not found in the trace represents a false-positive (FP). 
Similarly, a false-negative (FN) is a tracked BB that was missed by bcov. 
Both FPs and FNs represent errors in the reported coverage. 
Our evaluation method is conservative given the potential overapproximation in the CFG.
Also, we take into account the fact that \textsf{drcov} reports the heads of \textit{dynamic} BBs. 
This means that should A and B be consecutive BBs where A is fallthrough, i.e., does not end with a branch, 
\textsf{drcov} might only report the head of A.

Our results are shown in Table~\ref{tab:coverage-accuracy}.
They are based on running the test suites of subjects compiled with \textsf{gcc-7} in release build.
The results are representative of other build types.
The subjects are instrumented with {\bcov} and also run under the control of DR's \textsf{drcov}.
We list the average/maximum of TPs across all test processes. 
For example, the average number of TP BBs among 7,862 \textsf{llc} processes is 45,184.5, and the maximum is 90,952.
The average precision and recall across all subjects are 99.97\% and 99.95\% respectively.
This results in an average F-score of 99.86\%.
Our evaluation suggests that the reported coverage errors are practically negligible.
Nevertheless, there is still room for further improvement. 
Specifically, improving CFG precision and detour hosting can reduce FPs and FNs respectively. 



\section{Discussion}
\label{sec:discussion}
In this section, we discuss potential issues and limitations of {\bcov}.

\textbf{RISC ISA}.
Inserting detours is generally easy in RISC ISAs thanks to their fixed instruction size.
However, the addressing range can be significantly lower than the $\pm$2\,GB offered by x86-64.
Note that we patch each ELF module individually.
This means that we only need an addressing range that is large enough to reach our patch code segment from the original code.
For example, a range of 60MB would be sufficient for our largest subject.
AArch64 offers a detour range of $\pm$128\,MB  which can accommodate a large majority of binaries.
AArch32 offers just $\sim$32\,MB, in comparison.
In such a case, a single detour instruction might not be sufficient.
Additional options need to be investigated such as function relocation, literal pools, and modifications to linker scripts.

In addition, we update coverage data using a single pc-relative \texttt{mov} which has a memory operand with a 32-bit offset.
Generally, emulating the same functionality in RISC ISAs require more instructions and clobbering of registers.
However, saving and restoring the clobbered registers is not always necessary.
A liveness analysis can help us acquire registers with dead values. 
Similar analyses are already implemented in DBI tools.


\textbf{Limitations and threats to validity}.
The precision of the recovered CFG can affect the coverage reported by our tool.
While the implemented jump table and non-return analyses significantly increase CFG precision, 
they are still not perfect.
Our prototype might miss jump tables, albeit only in a few situations.
Also, while our experiments show that the non-return analysis in {\bcov} is comparable to IDA Pro, both tools face the challenge of \textit{may-return} functions.
Such functions might not return to their caller depending on their arguments.
Function \texttt{Perl\_\_force\_out\_malformed\_utf8\_message} in \texttt{perl} is particularly noteworthy.
In one binary, it is called 88 times (out of 89 total) with the argument \texttt{die\_here} set, i.e, will not return.
Developers can signal to the compiler that a particular call will not return using  \texttt{\_\_builtin\_unreachable()}.
Such hints are not available in the binary so we simply assume that all calls to may-return functions are returning.
Consequently, {\bcov} might spuriously report BBs following a may-return call as covered.

On another note, we believe that our subjects are representative of C/C++ software in Linux.
However, generalizing our results to other native languages and operating systems requires further investigation.
The simple mechanisms we use to implement detours and update coverage are also applicable to system software like kernel modules. 
However, special considerations might exist in such settings.
Finally, our approach cannot be directly applied to dynamic code, e.g., self-modifying code.


\section{Related Work}
\label{sec:related-work}

Instrumentation using trampolines is known for a long time.
It is typically used in restricted applications such as function interception~\cite{MicrosoftDetours}.
We systematically use trampolines at a fine granularity to instrument individual basic blocks.
Also, we are aggressive in exploiting padding bytes and hosting detours, which allows us to avoid relocating entire functions like in PEBIL~\cite{Laurenzano2010}.

Recently, several works considered static instrumentation via reassembly~\cite{Wang2015,RetroWriteSP2020} and recompilation~\cite{Anand:EuroSys2013}.
Both enable instrumentation code to be inlined in their recovered artifacts, namely, assembly and IR respectively. 
Therefore, they are orthogonal to our approach, in principle. 
However, code inlining means that relocated references need to be
fixed, e.g., in CFI records, which increases the engineering overhead. 
This can also be challenging to implement correctly since distinguishing references from scalars is an undecidable problem in general. 
In comparison, trampolines maintain reference stability which allowed us to seamlessly scale to large C/C++ binaries. 
Also, they make it easy to map analysis results back to the original binaries.

The analysis of jump tables was considered in several works.
A combination of pattern matching and data-flow analysis was proposed in~\cite{BenKhadra2016,Meng:ISSTA2016}. 
Cifuentes \etal~\cite{Cifuentes1999} use backward slicing to produce a slice, which they convert to a canonical IR expression before checking it against known jump table forms.
A custom value-set analysis using SMT solving was implemented in JTR~\cite{Cojocar2017a}.
It is applied after lifting instructions to LLVM IR.
In contrast, our approach, sliced microexecution, semantically reasons about jump tables without manual pattern matching. 
Also, it does not require lifting instructions to an IR. 
Such lifting is known to be error-prone~\cite{BinIR:ASE17} and can drastically slow down binary analyses~\cite{QSYMYun2018}.
Instead, we leverage the executable instruction semantics already available in off-the-shelf emulators.
Moreover, we move beyond mere recovery to jump table instrumentation.

Tikir \etal~\cite{Tikir2002} propose an approach for binary-level coverage analysis and use probe pruning to improve its efficiency.
It is the closest related work to ours. 
However, our approaches differ in several aspects. 
First, they focus on dynamic coverage analysis where binaries can be analyzed, patched, and potentially restored at runtime.
In contrast, our static instrumentation approach allows us to spend more time on optimizations.
Second, their work builds on Dyninst~\cite{DyninstWeb}, a generic binary instrumentation tool.
However, the generality of Dyninst comes at a considerable cost in terms of overhead and complexity. 
For example, it has multiple levels of trampolines.
In comparison, we focus on the bare minimum required for tracking code coverage. 
Consequently, {\bcov} provides better performance and transparency.
Finally, as acknowledged by the authors, the probe pruning
technique implemented in \bcov{} is more efficient than that of~\cite{Tikir2002}. 


\section{Conclusion}

In this work, we presented {\bcov}, a tool for binary-level coverage analysis.
We implement a trampoline-based instrumentation approach and demonstrate that it can be both efficient and transparent while scaling to large C/C++ programs.
However, this required an orchestrated effort where we leverage probe pruning, improve CFG precision, and cope with the instruction-size variability in x86-64 ISA.
Our tool statically instruments ELF binaries without compiler support. 
It largely avoids the need to modify the build system and, consequently, allows for high usability.
Also, we show that the produced coverage report is highly accurate, which can offer a valuable addition to the testing workflow.
We make our tool and dataset publicly available to foster further research in this area.

\section*{Acknowledgments}
We would like to thank the anonymous reviewers of  ESEC/FSE and USENIX Security for their insightful comments. 
Also, we would like to thank Fangrui Song for the helpful feedback.
Deeban Babu provided assistance with BAP and \texttt{angr} experiments. 
This research has been funded by the federal state of Rhineland-Palatinate, Germany.

\balance


\bibliographystyle{ACM-Reference-Format}
\bibliography{References}

\appendix

\end{document}